# Carboplatin binding to a model protein in non-NaCl conditions to eliminate partial conversion to cisplatin, and the use of different criteria to choose the resolution limit

**Simon W.M Tanley[a], Kay Diederichs[b], Loes M.J Kroon-Batenburg[c], Antoine M.M Schreurs[c] and John R Helliwell[a]***

[a] School of Chemistry, Faculty of Engineering and Physical Sciences, University Of Manchester, Brunswick Street, Manchester, M13 9PL, UK

[b] Department of Biology, University of Konstanz, D-78457 Konstanz, Germany

[c] Crystal and Structural Chemistry, Bijvoet Center for Biomolecular Research, Faculty of Science, Utrecht University, Padualaan 8, 3584 CH Utrecht, The Netherlands

Correspondence email: john.helliwell@manchester.ac.uk

## Synopsis

The binding of carboplatin to His-15 in hen egg white lysozyme in non-NaCl conditions is reported and thus removes the effect of carboplatin partially converting to cisplatin in the high NaCl conditions used previously. This study also investigates several criteria to determine the diffraction resolution limit.

## Abstract

Hen egg white lysozyme (HEWL) co-crystallisation conditions of carboplatin without sodium chloride (NaCl) have been utilised to eliminate partial conversion of carboplatin to cisplatin observed previously. Tetragonal HEWL crystals were successfully obtained in 65% MPD with 0.1M citric acid buffer at pH 4.0 including DMSO. The X-ray diffraction data resolution to be used for the model refinement was reviewed using several topical criteria together. The $CC_{1/2}$ criterion implemented in XDS led to data being significant to 2.0Å, compared to the data only being able to be processed to 3.0Å using the Bruker software package (SAINT). Then using paired protein model refinements and DPI values based on the FreeR value, the resolution limit was fine tuned to be 2.3Å. Interestingly this was compared with results from the EVAL software package which gave a resolution limit of 2.2Å solely using <I/sigI> crossing 2, but 2.8Å based on the Rmerge values (60%). The structural results showed that carboplatin bound to only the Nδ binding site of His-15 one week after crystal growth, whereas five weeks after crystal growth, two molecules of carboplatin are bound to the His-15 residue. In summary several new results have emerged: - firstly non-NaCl conditions showed a carboplatin molecule bound to His-15 of HEWL; secondly binding of one molecule of carboplatin was seen after one week of crystal growth and two molecules were bound after five weeks of crystal growth; and thirdly the use of several criteria to determine the diffraction resolution limit led to the successful use of data to higher resolution.

## 1. Introduction

Cisplatin and carboplatin are platinum anti-cancer drugs which have been used for a long time in the fight against cancer by targeting DNA. However, 90% of their reported binding is to plasma proteins (Fischer *et al* 2008). Thus, these drugs cause toxic side effects. Cisplatin is rapidly converted to toxic metabolites which cause nephrotoxic effects (Zhang, 1996; Huliciak *et al*, 2012), whereas carboplatin is less toxic due to the addition of the cyclobutanedicarboxylate moiety, which has a slower rate of conversion. Carboplatin can therefore be tolerated by patients at higher doses compared to cisplatin (Kostova, 2006). Previous crystallographic studies of cisplatin and carboplatin with hen egg white lysozyme (HEWL), a model protein, have shown binding of two molecules to its His-15 residue in dimethyl sulfoxide (DMSO) media (Tanley *et al*, 2012a; 2012b; Helliwell & Tanley 2013) and for cisplatin even after prolonged exposure in an aqueous medium (Tanley *et al*, 2012b). Subsequently, through public archiving of our raw diffraction images at Utrecht University (Tanley *et al*, 2013, http://rawdata.chem.uu.nl/#0001;) and now also mirrored at the Tardis Raw data archive in Australia (http://vera183.its.monash.edu.au/experiment/view/40/), a collaboration was set up with one of the authors of this article (KD) who downloaded and re-processed the diffraction images, measured in Manchester, with the XDS software package (Kabsch, 2010) to compare with our previous results. Reviewing these results along with the previous publication of the carboplatin bound structures in DMSO media studied at cryo and room temperatures (Tanley *et al*, 2012a; Tanley *et al*, 2012b) (PDB id's 4dd7, 4dd9 and 4g4c) it was noted that the anomalous difference electron density peaks within the carboplatin binding sites weren't only for the Pt centres (Tanley *et al,* 2014, Accepted for publication in the *Journal of Synchrotron Radiation*). Thus, this suggested that with the high sodium chloride (NaCl) concentrations used in our crystallisation conditions (Tanley *et al*, 2012a; 2012b; Helliwell & Tanley, 2013) the carboplatin could be partially converted to cisplatin, with the extra anomalous difference density thereby being attributed to the Cl atoms of cisplatin. This partial conversion of carboplatin has been seen previously in solution (Gust & Schnurr, 1999). Due to these new findings, the binding sites could thus contain a mixture of carboplatin and cisplatin rather than just the carboplatin molecule, which is what we initially thought was solely bound (Tanley *et al*, 2014, Accepted for publication in the *Journal of Synchrotron Radiation*).

Based on these new findings, this study focuses on crystallising HEWL with carboplatin in non-NaCl conditions to remove the possibility of carboplatin converting to cisplatin. Crystals were successfully grown without NaCl and the results are presented here. The raw diffraction data images for these MPD chemical medium grown crystals were processed using the Bruker software program (SAINT) as well as EVAL (Schreurs *et al*, 2010) and XDS (Kabsch, 2010).

Determining the high resolution cut off limit for X-ray crystallographic data is a major discussion point (Diederichs & Karplus, 2013; Evans & Murshudov, 2013), with Rmerge values being used for many years (Arndt *et al*, 1968) and in recent years supplanted by monitoring where <I/sigI> crosses a value of 2.0. Rmerge measures the relative spread of n independent measurements of the intensity of a reflection around their average. A relatively new criterion has been implemented, $CC_{1/2}$, which is based on the Pearson correlation coefficient of two half datasets (Karplus & Diederichs, 2012; Evans, 2011). Our study looks to determine the high resolution cut off of our newly processed datasets using these multiple criteria and, notably, to see if one can use data to a higher resolution than indicated from <I/sigI> crossing 2.0.

Several new results have emerged: - Firstly non-NaCl conditions show a carboplatin molecule bound to His-15 of HEWL; secondly binding of one molecule of carboplatin was seen after one week of crystal growth and two molecules were bound after five weeks of crystal growth; and thirdly the use of several criteria combined allowed us to more carefully assess the diffraction resolution limit than hitherto.

## 2. Methods

### 2.1. Crystallisation conditions.

20mg of HEWL (0.6mM) was dissolved in 1ml distilled water. 1.4mg of carboplatin (1.8mM) was added in a 3-fold molar excess over that of the protein, along with 75µl DMSO and mixed until all the carboplatin had dissolved. 48 non-NaCl crystal screens from Hampton were set up; these comprised 2µl protein/carboplatin/DMSO solution aliquots each mixed with 2µl reservoir solution and were set up in hanging drop crystallisations with 1ml of reservoir solution. Crystallisation trays were left at room temperature and crystals grew after ~ 4 weeks in the condition: 65% MPD with 0.1M citric acid buffer at pH 4.0. The citrate ions removing a proton from the histidine nitrogen(s) would start immediately after mixing the solutions.

### 2.2. X-ray diffraction data collection, protein structure solution and model refinement

A 0.3mm sized crystal grown in the 65% MPD conditions was mounted into a loop with silicon oil used as the cryoprotectant and X-ray diffraction data were measured one week after crystal growth (crystals took ~ 4 weeks to grow). The crystal was mounted on a Bruker APEXII home source diffractometer and data collection was carried out at 110K with a data collection strategy used to gain the most information possible. A second crystal (0.1mm sized), five weeks after initial crystal growth, was mounted into a loop again with silicon oil used as the cryoprotectant. The crystal was again mounted on the same Bruker APEXII home source diffractometer and X-ray diffraction data collection was carried out at 127K with a data collection strategy again used to gain the most information possible. The two crystal datasets were processed using the Bruker software package (SAINT) as well as EVAL (Schreurs *et al*, 2010) and XDS (Kabsch, 2010) software packages. The structures were solved using molecular replacement with PHASER (McCoy *et al*, 2007) and restrained refinement with CCP4i REFMAC5 (Vagin *et al*, 2004), using the reported lysozyme structure 2w1y as molecular search model (Cianci *et al* 2008). Model building, adjustment and refinement were carried out respectively using the COOT (Emsley & Cowtan, 2004) molecular graphics programme and REFMAC5 (Vagin, 2004) in CCP4i. Ligand binding occupancies were calculated using SHELXTL (Sheldrick, 2008). Crystallographic and refinement parameters for crystals 1 and 2 processed using the SAINT, EVAL and XDS software packages are summarized in Table 1.

**Table 1**

X-ray crystallographic data and final protein model refinement statistics for crystals 1 and 2, processed by 3 different software packages.

| | Crystal 1 | | | Crystal 2 | | |
|---|---|---|---|---|---|---|
| | Bruker (SAINT) | EVAL | XDS | Bruker (SAINT) | EVAL | XDS |
| PDB id | Not deposited | 4lt0 | 4lt1 | Not deposited | 4lt2 | 4lt3 |
| Data collection temperature (K) | 110 | 110 | 110 | 127 | 127 | 127 |
| **Data reduction** Space group | $P4_32_12$ | $P4_32_12$ | $P4_32_12$ | $P4_32_12$ | $P4_32_12$ | $P4_32_12$ |
| Unit cell parameters (Å)/(°) | a=b= 76.97 | a=b= 76.77 | a=b= 76.82 | a=b= 77.12 | a=b= 77.29 | a=b= 77.12 |
| | c= 36.28 | c= 36.36 | c= 36.47 | c= 36.54 | c= 36.63 | c= 36.56 |
| | α=β=γ= 90.0 | α=β=γ= 90.0 | α=β=γ= 90.0 | α=β=γ= 90.0 | α=β=γ= 90.0 | α=β=γ= 90.0 |
| Protein molecular Mass | 14700 | 14700 | 14700 | 14700 | 14700 | 14700 |
| Molecules per asymmetric unit | 1 | 1 | 1 | 1 | 1 | 1 |
| Detector to crystal distance (mm) | 40.20 | 40.20 | 40.20 | 40.37 | 40.37 | 40.37 |
| Observed reflections | 58539 | 202135 | 231303 | 187183 | 169869 | 206299 |
| Unique reflections | 4211 | 7770 | 7765 | 6323 | 7901 | 7850 |
| Resolution (Å) (last shell) | 36.28 – 3.00 (3.10 – 3.00) | 19.18 – 2.00 (2.03 – 2.00) | 39.41 – 2.00 (2.05 – 2.00) | 36.54 – 2.16 (2.19-2.16) | 25.15 – 2.00 (2.04 – 2.00) | 39.41 – 2.00 (2.05 – 2.00) |
| Completeness (%) | 99.8 (99.1) | 99.8 (99.8) | 99.7 (99.8) | 100 (100) | 98.9 (93.7) | 99.4 (32.4) |
| Rmerge (%) | 0.255 (0.563) | 0.340 (2.71) | 0.422 (3.935) | 0.227 (0.738) | 0.193 (1.41) | 0.248 (2.27) |
| $(I/\sigma(I))$ | 11.4 (2.8) | 9.2 (0.9)\$a | 8.9 (0.5)\$b | 14.4 (2.01) | 14.2 (1.1)\$c | 14.6 (0.5)\$d |
| Multiplicity | 23.9 (9.8) | 26.1 (12.3) | 29.7 (9.8) | 29.6 (13.3) | 21.6 (5.2) | 26.3 (6.0) |
| $CC_{1/2}$ | - | 0.38 | 0.16 | - | 0.51 | 0.32 |
| Cruickshank DPI (Å) | 0.59 | 0.25* | 0.27** | 0.24 | 0.21 | 0.19 |
| Average B factor (Å$^2$) | 22.3 | 26.8* | 24.4** | 25.5 | 22.8 | 27.6 |
| **Refinement** | | | | | | |
| R factor/ R free | 20.2/28.9 | 22.3/28.3* | 18.5/24.8** | 18.4/26.0 | 20.7/26.5 | 19.5/25.7 |
| R factor all | 20.6 | 22.6* | 18.8** | 18.8 | 21.0 | 19.8 |
| RMSD bonds (Å)/ Angles (°) | 0.011/ 1.634 | 0.016/1.888* | 0.014/1.644** | 0.015/1.691 | 0.016/1.815 | 0.014/1.650 |
| **Ramachandran values (%)** | | | | | | |
| Most favoured | 92.1 | 96.1 | 96.8 | 96.8 | 96.1 | 97.6 |
| Additional allowed | 7.9 | 3.9 | 3.2 | 3.2 | 3.9 | 2.4 |
| Disallowed | 0 | 0 | 0 | 0 | 0 | 0 |

\$a <I/sigI> = 2.4 at 2.2Å, <I/sigI> = 1.7 at 2.15Å

\$b <I/sigI> = 2.0 at 2.24Å

\$c <I/sigI> = 2.3 at 2.07Å, <I/sigI> = 1.7 at 2.03Å

\$d <I/sigI> = 2.17 at 2.17Å, <I/sigI> = 1.77 at 2.11Å

* Final refinement statistics to 2.1Å for crystal 1 processed by EVAL, whereas the data reduction statistics are to 2.0Å; see also Table 2 a and b.

** Final refinement statistics to 2.3Å for crystal 1 processed by XDS, whereas the data reduction statistics are to 2.0Å; see also Table 2 a and b.

# 3. Results

## 3.1 Data processing

Data processing of the raw diffraction images was carried out using SAINT (Bruker internal software), the XDS package (Kabsch, 1988) and EVAL (Schreurs *et al*, 2010). The use of these three different processing programs for crystal 1 in this case was vital, with the SAINT (Bruker) software processing the data to 3.0Å resolution (<I/sigI> = 2.8 to 3.0Å), while processing to higher resolutions failed. However using XDS, we found that based on the $CC_{1/2}$ criterion (Karplus & Diederichs, 2012) there was information in the data up to 2.0Å resolution. Using EVAL, the data would have been cut at 2.2Å based on where <I/sigI> cross 2, or 2.8Å based on the Rmerge values of this dataset exceeding 60%. However, as the $CC_{1/2}$ values from the XDS processing showed data out to 2.0Å resolution still carried information, the raw diffraction data images were also processed to 2.0Å resolution using EVAL. Due to the differences in the resolution limit chosen between SAINT, EVAL and XDS, separate pairwise refinements were used for the XDS and EVAL data to determine which resolution was optimal (Table 2), along with the DPI values based on the FreeR value (Figure 1), with 2.3Å being chosen via either criterion for the XDS processed raw diffraction data images and 2.1Å for the EVAL processed raw diffraction data images. The best final model obtained from the XDS and EVAL data, respectively, was used for calculation of R-values against both datasets, allowing unambiguous identification of the best data and model (Table 3), with the EVAL processed data and model to 2.1Å resolution being marginally better than the XDS processed data and model to 2.3Å.

**Table 2**   Pairwise refinements (Karplus & Diederichs, 2012) showing R/Rfree of initial models refined at given resolution (top row), against data as given in left column. (a) XDS data and (b) is the EVAL data. Bold numbers indicate improvement of Rfree compared to refinement at lower resolution. The resolution which yields the consistently best model is highlighted in yellow; this resolution cut-off is hereby determined as optimal for these data, the given protein model, the refinement strategy and refinement program.

| (a) | | **Resolution that model was refined at** | | | | | | |
|---|---|---|---|---|---|---|---|---|
| | | 3.0A | 2.5 Å | 2.4 Å | 2.3 Å | 2.2 Å | 2.1 Å | 2.0 Å |
| | 3.0 Å | 16.4/28.7 | 16.1/28.4 | 16.2/28.8 | 16.3/28.5 | 16.3/28.4 | 16.3/28.4 | **16.3/28.1** |
| | 2.5 Å | | 17.7/25.6 | 17.6/26.2 | **17.7/25.4** | 17.6/25.4 | 17.6/25.5 | **17.6/25.4** |
| **Resolution used for R-value calculation** | 2.4 Å | | | 17.9/25.7 | 17.9/25.7 | **17.9/25.6** | **17.9/25.6** | 17.9/25.5 |
| | 2.3 Å | | | | <mark>18.5/24.8</mark> | **18.6/24.7** | **18.5/24.7** | 18.8/24.8 |
| | 2.2 Å | | | | | 18.8/25.9 | 18.8/25.9 | 18.8/25.9 |
| | 2.1 Å | | | | | | 19.7/27.2 | 19.7/27.3 |
| | 2.0 Å | | | | | | | 21.3/28.4 |

| (b) | **Resolution that model was refined at** |
|---|---|

| | | 3.0A | 2.5 Å | 2.3 Å | 2.15 Å | 2.1 Å | 2.05 Å | 2.0 Å |
|---|---|---|---|---|---|---|---|---|
| | 3.0 Å | 19.2/28.7 | **19.0/27.4** | 18.9/28.0 | 18.9/27.7 | **18.6/27.4** | 18.9/27.9 | 18.9/27.6 |
| | 2.5 Å | | 20.7/29.9 | 20.7/30.1 | 20.7/29.8 | **20.5/29.4** | 20.7/29.6 | 20.2/29.6 |
| Resolution used for R-value calculation | 2.3 Å | | | 21.1/28.3 | 21.3/28.4 | 21.3/28.3 | 21.3/28.3 | **21.3/28.2** |
| | 2.15Å | | | | 22.1/28.3 | **22.1/28.2** | 22.2/28.3 | 22.2/28.3 |
| | 2.1 Å | | | | | ==**22.3/28.3**== | 22.4/28.3 | **22.3/28.3** |
| | 2.05Å | | | | | | 22.5/28.9 | 22.5/28.9 |
| | 2.0 Å | | | | | | | 23.1/29.2 |

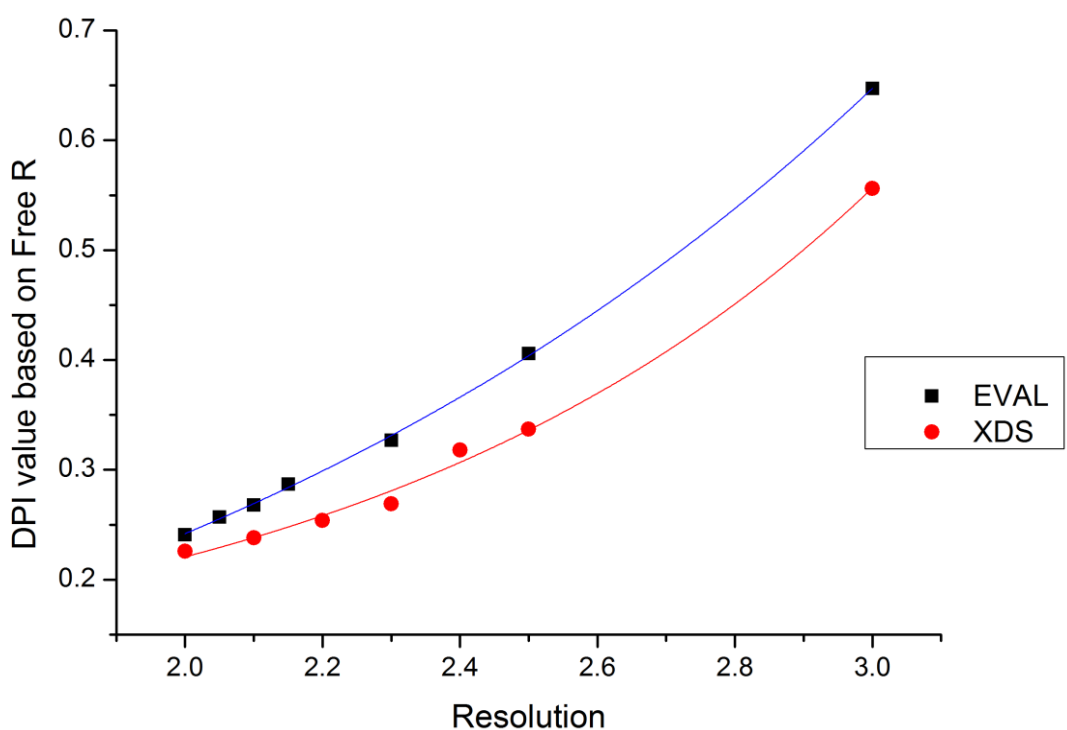

**Figure 1** DPI value based on the FreeR factor for each refined model at different resolutions for both the XDS and EVAL processed raw diffraction data images. The FreeR set was kept the same for both the XDS and EVAL processed datasets. The curves are the line of best fit through the data points using a power 5/2 fit (Our data follows the same trends as in Blow, 2002) in OriginPro 8.5.1 with the line showing excellent agreement with the data points .

**Table 3** The pairwise comparison technique applied to the XDS and EVAL data and final refined models. The best model obtained from the XDS and EVAL data, respectively, was used for calculation of R-values against both datasets, allowing unambiguous identification of the best data and model. Highlighting in yellow shows the EVAL final model and data at 2.1Å resolution has lower Rfree compared to the XDS model against the EVAL data. Also, the EVAL final model has the same Rfree as the XDS final model against the XDS data.

| | | Resolution that model was refined at | |
|---|---|---|---|
| | | **2.1 Å EVAL** | **2.3 Å XDS** |
| **Resolution used for R-** | **2.1 Å EVAL** | 22.3/28.3 | 24.4/29.5 |
| **value calculation** | **2.3 Å XDS** | 20.9/24.8 | 18.5/24.8 |

### 3.2 Carboplatin binding to His-15

For crystal 1, collected one week after crystal growth shows carboplatin bound to the Nδ binding site of His-15 at 3.0Å from the SAINT (Bruker) internal software package (Figure 2a), 2.3Å from the XDS (Figure 2b) software package and 2.1Å from the EVAL (Figure 2c) software package. The models at 2.3Å and 2.1Å (Figures 2b and c) show slightly more detail in the 2Fo-Fc map at the Nδ binding site, as more detail for the atoms bound to the Pt centre can be seen compared to just the Pt atom being visible in the density to 3.0Å. Binding of the carboplatin molecule to the Nε atom however, is not seen. The anomalous difference density is however weak as determined from all three software packages, with the highest peak being 5.1σ at the Pt position, when the data was cut to 4.0Å resolution (Table 4).

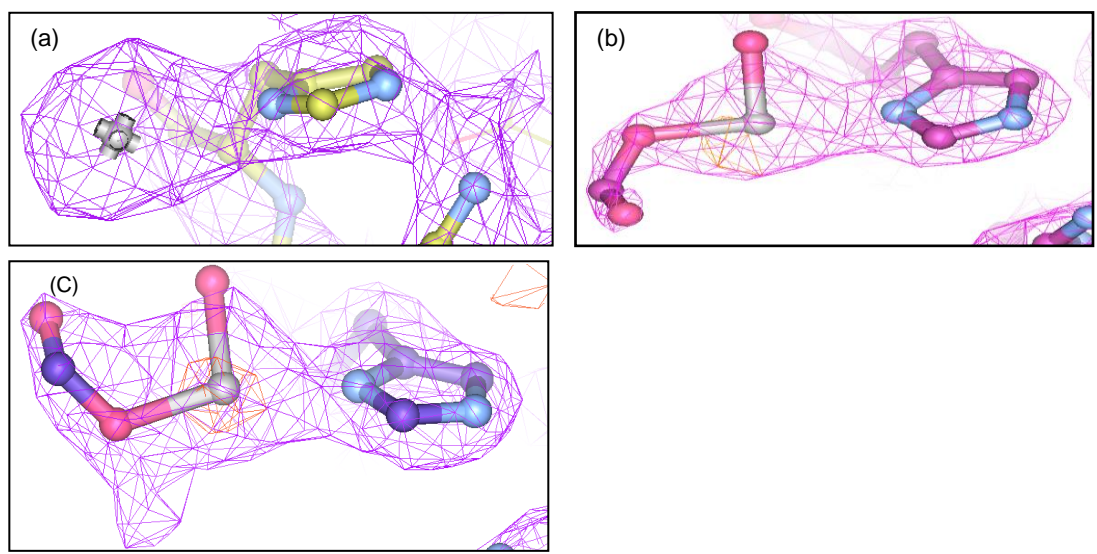

**Figure 2**  Binding of carboplatin to the Nδ atom of His-15 from crystal 1 at (a) 3.0Å from the SAINT (Bruker) processed data (b) 2.3Å from the XDS processed data and (c) 2.1Å from the EVAL processed data. 2Fo-Fc maps are shown at the 1.2σ cut off level. Anomalous difference electron density (orange) maps are shown at the 3.0σ cut off level. No anomalous difference electron density was present in the Bruker dataset map (a).

For crystal 2 (Figure 3), collected five weeks after crystal growth shows binding of two molecules of carboplatin, one each bound to the Nδ and the Nε atom of His-15. The same as for crystal 1, the anomalous difference density for the Pt atoms is weak, with the highest peak being 4.6σ for the Nδ binding site and 5σ for the Nε binding site (Table 4).

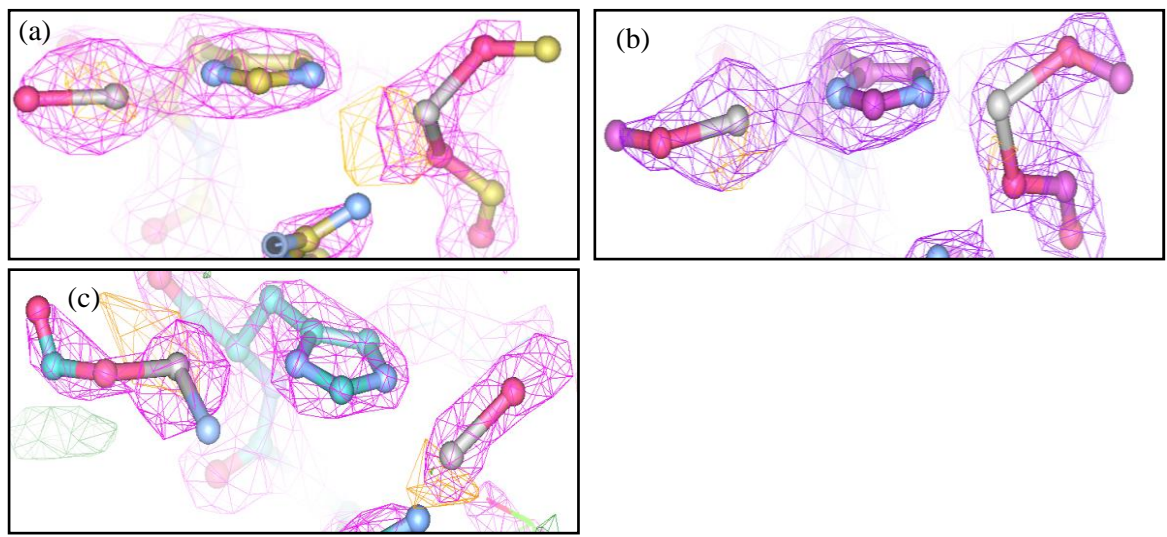

**Figure 3** Binding of carboplatin to the Nδ and Nε atoms of His-15 for crystal 2 at (a) 2.16Å from the SAINT (Bruker) processed data (b) 2.0Å from the XDS processed data and (c) 2.0Å from the EVAL processed data. 2Fo-Fc maps are shown at the 1.2σ cut off level. Anomalous difference electron density (orange) maps are shown at the 3σ cut off level.

**Table 4**   Anomalous difference density peak heights for the Pt position at both the Nδ and Nε binding sites for crystals 1 and 2 along with the occupancy values of the Pt atoms, calculated using SHELX (Sheldrick, 2008).

|  | Nδ | | Nε | |
|---|---|---|---|---|
|  | Anomalous peak height (σ) | Pt occupancy (%) | Anomalous peak height (σ) | Pt occupancy (%) |
| Crystal_1_Bruker | 0 | 47 | - | - |
| Crystal_1_EVAL | 5.1 | 39 | - | - |
| Crystal_1_XDS | 3.4 | 29 | - | - |
| Crystal_2_Bruker | 4.1 | 42 | 4.6 | 16 |
| Crsytal_2_EVAL | 4.6 | 22 | 5.0 | 31 |
| Crystal_2_XDS | 4.4 | 23 | 3.1 | 38 |

## 4.0 Discussion

### 4.1 Carboplatin binding to His-15 in non-NaCl conditions

Crystals were grown in 65% MPD with 0.1M citric acid buffer at pH 4.0 after ~4 weeks, and one week after crystal growth one molecule of carboplatin was seen bound to the Nδ atom of His-15 (Figure 2). However, no binding was seen to the Nε binding site. At the low pH used (pH 4.0) in these crystallisation conditions, the His-15 residue is assumed to be a protonated histidine, which means that both N-hydrogen atoms would need to be removed before binding to the Pt centre can be achieved at these histidine nitrogens. In the earlier studies we have observed binding of two molecules of cisplatin and carboplatin to His-15 in HEWL (Tanley *et al*, 2012a; 2012b; Helliwell & Tanley, 2013), one to each nitrogen, and was determined to be due to the histidine residue being an imidazolyl anion which is formed by the removal of the usual N-hydrogen atoms, which could be brought about by either the Cl or acetate ions in the co-crystallisation conditions used in those studies (Tanley *et al*, 2012a). However, as the MPD crystallisation conditions do not contain either Cl or acetate ions, this leaves the citrate ions in the crystallisation mixture deemed to be able to extract an N-hydrogen atom. As these crystals took 4 weeks to grow and then were measured both one week and five weeks after crystal growth, the chemistry of the citrate ions removing a proton from the histidine nitrogen(s) would start immediately after the conditions were set up. Thus, the equilibrium in de-protonation of the His N atoms and binding to cisplatin is reached slower compared to the previous crystallisation conditions which include the Cl and acetate ions as only one binding site for carboplatin is seen one week after the crystals have grown (take ~4weeks to grow), compared to the two binding sites we previously saw after around 8 days of crystallisation (Tanley *et al*, 2012a; 2012b; Helliwell & Tanley, 2013). For the case of our second crystal, which was measured five weeks after crystal growth, binding of two molecules of carboplatin to the His-15 residue is seen (Figure 3). Thus, the extraction of the second N-hydrogen atom took longer due to the relatively low concentration of citrate ions in these MPD medium crystallisation conditions. The occupancy values for the Pt atoms in this study (mean occupancy of ~30% for both the Nδ and Nε binding sites, Table 4) are lower compared to the occupancies seen when the crystals were grown in the high NaCl conditions used previously (~70% for the Nδ binding site and ~50% for the Nε binding site) (Tanley *et al*, 2012a; 2012b; Helliwell & Tanley, 2013).

## 4.2 Determining the diffraction resolution of the data

As well as now seeing just carboplatin bound in the MPD conditions, this study also looked at the use of combining several criteria to determine the diffraction resolution limit of these data, and indeed which led to the successful use of data to higher resolution. The $CC_{1/2}$ criterion in particular (Karplus & Diederichs, 2012; Diederichs & Karplus, 2013) for determining the resolution limit for a given dataset meant that for crystal 1, data to a higher resolution was used compared to using the traditional statistics to determine the resolution limit of the data (Rmerge values or where <I/sigI> crosses 2). Once the data were processed and refined, the pairwise refinements technique (Karplus & Diederichs, 2012; Diederichs & Karplus, 2013) was used to determine which resolution was best for these data. In this case, using $CC_{1/2}$ (Table 1), the pairwise refinement technique (Table 2) and the DPI value (Cruickshank, 1999) of the refined model using the Rfree factor (Figure 1), led to 2.3Å resolution being determined for the XDS processed diffraction data images and 2.1Å resolution for the EVAL processed diffraction data images for crystal one, compared to 3.0Å from SAINT (the Bruker internal software package), with attempts to process the data to higher resolution using SAINT failing to produce any results. In Table 2, the highlighting in bold represents the improvement in R/Rfree values against the resolution used to refine the model and for the XDS case the refined model at 2.3Å resolution has the lowest R/Rfree gap (6.3) of all the models, meaning that for the XDS processed data, 2.3Å resolution is most probably the optimum high resolution cut off to use for this dataset, but is not clear cut. However, for the EVAL processed data, it is clear that 2.1Å resolution is the optimum resolution to use for this dataset. The graphs of the DPI values based on the Rfree value (Figure 2) agrees with these results from the pair-wise refinement technique as at these resolutions, one can see a flattening off of the curve agreeing with the results in Blow, 2002. Plotting the natural logs of the DPI values and the resolution (Supplementary Figure 1), EVAL shows a power law of 2.4 and XDS 2.3, ie very close to the theory value of 2.5 (Blow, 2002) which shows that the DPI is proportional to the power 5/2. Interestingly, the XDS processed dataset to 2.3Å resolution has a very similar DPI value to the EVAL processed dataset to 2.1Å resolution. If the $CC_{1/2}$ criterion had not been available, this dataset would have been processed by EVAL to 2.8Å based on the Rmerge values being higher than 60% and using <I/sigI> crossing 2 this was 2.2Å but had an Rmerge value of 140%. Thus, processing these data with multiple software packages with these different criteria to determine the overall resolution limit led to use of higher resolution.

The best model obtained from the XDS and EVAL data, respectively, was used for calculation of R-values against both datasets, allowing unambiguous identification of the best data and model (Table 3). From these pair-wise refinements, it is determined that the EVAL processed data to 2.1Å and the final refined model is marginally better than the XDS processed data to 2.3Å and its final refined model. The reason why the R-values (Rwork/Rfree) of the best model against the XDS data are about 2% lower than the best model against the EVAL data is not known to us, and this question is currently under investigation.

The higher high-resolution limit of the EVAL data compared with the XDS data is consistent with the CC1/2 values reported by EVAL at 2.1Å resolution (80%) and XDS at 2.3Å resolution (75%) respectively.

As a further comparison for crystal 2, this dataset was processed to 2.16Å by SAINT, which was where the <I/sigI> value crossed 2, but with $CC_{1/2}$ implemented in both XDS and EVAL, this dataset was processable to 2.0Å; the edge of the detector, which therefore limited the high resolution for crystal 2.

Comparing the XDS and EVAL unscaled and scaled data for crystal 1 leads to a number of conclusions about this particular dataset; (i) the statistics of the EVAL data including the $CC_{1/2}$ criterion are slightly better than XDS. (ii) The difference between XDS and EVAL is that large intensity reflections are larger in EVAL than XDS. (Supplementary Figure 2a) (iii) The standard deviations of individual reflections in EVAL are larger than in XDS (Supplementary Figure 2b). The difference is largely caused by SADABS and XSCALE applying different error models. SADABS relies heavily on internal standard deviations (iv) The refinement results are better (~2% lower Rfactors) for XDS. (v) The correlation between the F's from XDS and EVAL is very good, except for the lowest resolution, where some reflections are partially screened by the beam stop, the definition of the beam stop being slightly different in EVAL from XDS.

## 5.0 Conclusions

The study outlined here shows carboplatin bound to the Nδ and Nε atoms of His-15 from HEWL in non-NaCl conditions, thus removing the possibility of carboplatin partially converting to cisplatin in high salt conditions seen previously (Tanley et al 2014, Accepted for publication in the *Journal of Synchrotron Radiation*).

This study also looked at the use of several criteria to determine the resolution range of the datasets. Using $CC_{1/2}$, the pairwise refinement technique and the DPI value based on FreeR, 2.1Å was confirmed as the best resolution limit to use for the EVAL processed raw diffraction images.

## Acknowledgements

We are grateful for research support from the Universities of Konstanz, Manchester and Utrecht. ST is funded under an EPSRC PhD Research Studentship. We are grateful to Dr Ed Pozharski, University of Maryland for valuable discussions.

## Supplementary Materials

**Figure S1** Natural log graph of the DPI value based on Rfree against the resolution limit of the refined model for both the XDS and EVAL processed datasets. One would expect a power law of 2.5 based on equation 9 of Blow, 2002. The EVAL data has a power law of 2.4, and XDS 2.3, but both datasets follow the same trend (Figure 1), similar to the results in Blow, 2002.

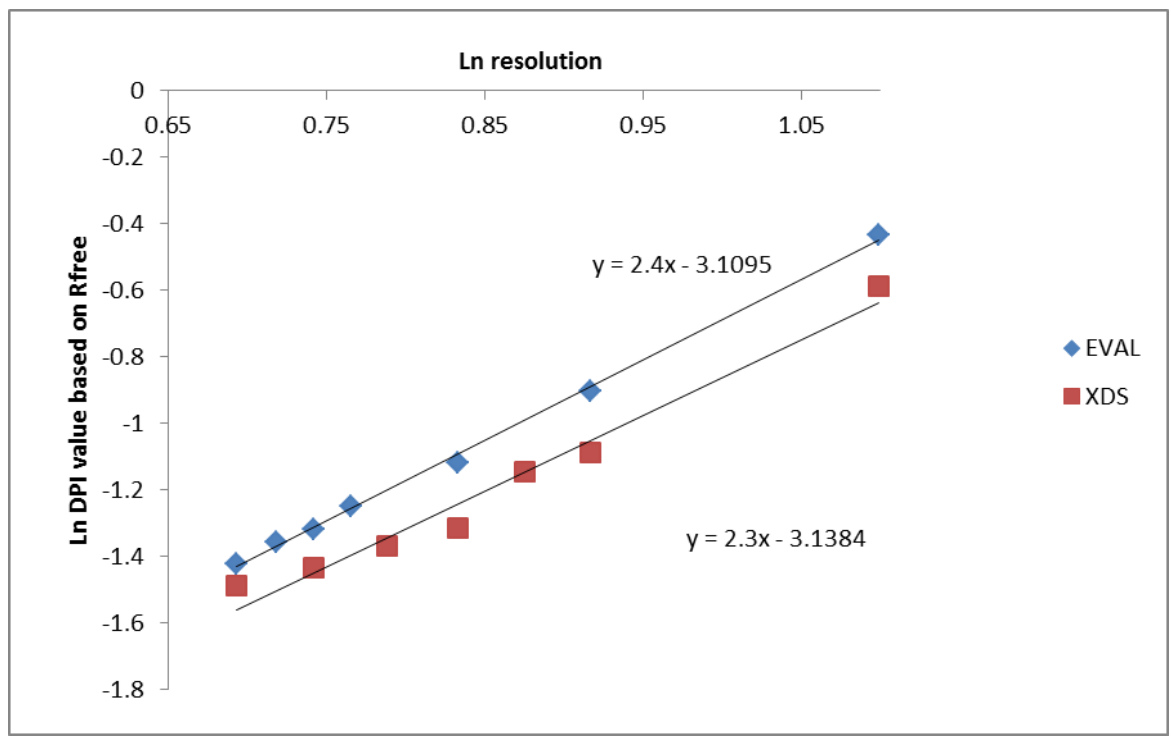

**Figure S2** Comparison between XDS and EVAL for the data of crystal 1. (a) $Log_{10}$ (intensity) of reflections, which are larger in EVAL (horizontal) than XDS (vertical) and (b) I/σ comparison of EVAL vs XDS data. The error model in SADABS, used for the EVAL data, is largely responsible for the observed differences in I/σ.

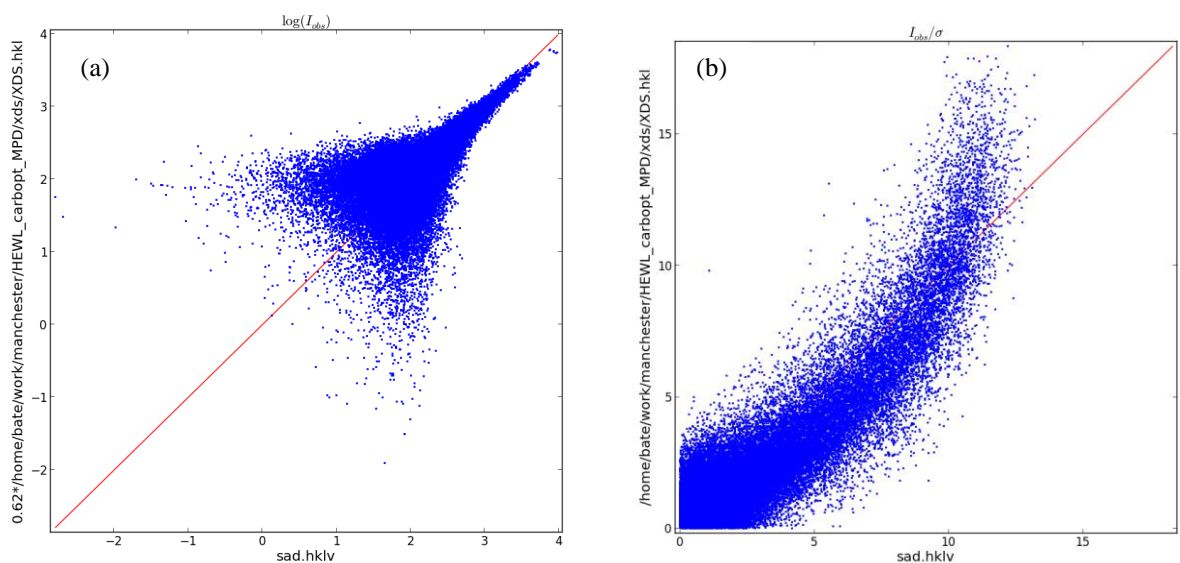

## References

Arndt, U.W, Crowther, R.A & Mallet, J.F.W (1968). *J Phys. E Sci Instrum.* **1**:510-516

Blow, D.M (2002). *Acta Cryst D***58**:792-797

Cianci, M., Helliwell, J. R. & Suzuki, A. (2008). *Acta Cryst*. *D***64:**1196–1209.

Cruickshank, D. W. J. (1999). *Acta Cryst. D* **55**: 583–601.

Diederichs, K & Karplus P.A. (2013) *Acta Cryst D* **69**:1215-1222

Emsley, P. & Cowtan, K. (2004). *Acta Cryst*. *D***60**, 2126–2132.

Evans, P.R. (2011) *Acta Cryst D* **67**:282-292

Evans, P.R. & Murshudov, G.N. (2013) *Acta Cryst D***69**:1204-1214

Gust, R. & Schnurr, B. (1999). *Monatsh. Chem.* **130**, 637–644.

Fischer, S. J., Benson, L. M., Fauq, A., Naylor, S. & Windebank, A. J. (2008). *Neurotoxicology*, **29**, 444–452.

Helliwell, J.R & Tanley, S. W.M (2013) *Acta Cryst D* **69**:121-125

Huličiak M, Vacek J, Sebela M, Orolinová E, Znaleziona J, Havlíková M, *et al.* (2012). *Biochemical pharmacology* **83**: 1507–1

Kabsch, W. (2010) *Acta Cryst D* **66:**125-132

Karplus P.A., Diederichs, K. (2012). *Science* **336**, 1030-1033

Kostova, I. (2006). *Recent Pat. Anticancer Drug. Discov*. **1**, 1–22.

McCoy, A. J., Grosse-Kunstleve, R. W., Adams, P. D., Winn, M. D.,Storoni, L. C. & Read, R. J. (2007). *J. Appl. Cryst*. **40**, 658–674.

Schreurs, A. M. M., Xian, X. & Kroon-Batenburg, L. M. J. (2010). *J.Appl. Cryst*. **43**, 70–82.

Sheldrick, G.M. (2008). *Acta Cryst*. A**64**, 112-122

Tanley S.W.M, Schreurs AMM, Kroon-Batenburg LMJ, Meredith J, Prendergast R, Walsh D, Bryant P, Levy C, Helliwell JR (2012a) *Acta Cryst D* **68**:601-612

Tanley, S.W.M, Schreurs, A. M. M.,  Kroon-Batenburg, L. M. J, and Helliwell, J. R. (2012b) *Acta Cryst*. *F***68**, 1300–1306.

Tanley, S.W.M, Schreurs, A. M. M., Helliwell, J. R. and  Kroon-Batenburg, L. M. J.  (2013) *J. Appl. Cryst.* **46**, 108-119

Tanley, S.W.M, Diederichs, K, Kroon-Batenburg, L.M.J, Schreurs, A.M.M and Helliwell, J.R (2014). Accepted for publication in the *Journal of synchrotron radiation ISDSB2013* conference paper.

Vagin, A. & Teplyakov, A. (2010). *Acta Cryst*. D**66**, 22–25.

Zhang J. (1996). C. *Toxicology Letters* **89**: 11–17.